\documentclass{PoS}

\usepackage{amsmath}
\usepackage{amssymb}
\usepackage{graphicx}

\title{Classification and Generalization of Minimal-doubling actions}

\ShortTitle{Classification of Minimal-Doubling Fermions}

\author{\speaker{Tatsuhiro Misumi}\thanks{Supported 
        by Grand-in-Aid for the Japan Society for 
        Promotion of Science (JSPS) Research Fellow No.\ 21-1226.}\\
        Yukawa Institute for Theoretical Physics, Kyoto University\\
        E-mail: \email{misumi@yukawa.kyoto-u.ac.jp}}

\author{Michael Creutz \thanks{
This manuscript has been authored by employees of
 Brookhaven Science Associates, LLC under Contract
 No. DE-AC02-98CH10886 with the U.S. Department of Energy. The
 publisher by accepting the manuscript for publication acknowledges
 that the United States Government retains a non-exclusive, paid-up,
 irrevocable, world-wide license to publish or reproduce the published
 form of this manuscript, or allow others to do so, for United States
 Government purposes.
        }\\
        Brookhaven National Laboratory\\
        E-mail: \email{creutz@bnl.gov}}

\author{Taro Kimura\\
        Department of Basic Science, University of Tokyo\\
        E-mail: \email{kimura@dice.c.u-tokyo.ac.jp}}

\abstract{We propose a method to control the number of species of lattice
fermions, which yields new classes of minimally doubled lattice
fermions with one exact chiral symmetry and exact
locality. We classify all the known minimally doubled fermions into two
types based on the locations of the propagator poles in the Brillouin
zone. We also study higher-dimensional extension of them and show 
it tends to be more difficult to realize minimal-doubling in higher dimensions.}

\FullConference{The XXVIII International Symposium on Lattice Field Theory, Lattice2010\\
		June 14-19, 2010\\
		Villasimius, Italy}

\begin{document}

\section{Introduction}
\label{sec:Intro}
The doubling problem of fermions has been one of the notorious obstacles 
to QCD simulations.    
By now several fermion constructions to bypass it
have been developed, although all of them have their individual shortcomings.
There is also an alternative possibility to settle the problem.   
More than 20 years ago,
Karsten and Wilczek proposed the minimally doubled action \cite{KW},
where only two species emerge.  
Recently one of the authors of the present paper
\cite{BC} has also proposed a two-parameter class
of fermion actions.  
These minimally doubled fermions all
possess one exact chiral symmetry and exact locality.  As such they
should be faster for simulation, at least for two-flavor QCD, than
other chirally symmetric lattice fermions.
In this paper we propose a systematic method to reduce the number of
doublers, which we term twisted-ordering method \cite{CM}.  
In this way we obtain new classes of minimally-doubled fermions. 
We classify all the known minimally doubled actions into two types.  
By this classification we can derive several further minimally
doubled actions deductively.
We also study higher-dimensional extension of them and find 
the parameter range for minimal-doubling gets narrower with the dimension 
in most cases \cite{KM}.


\section{Twisted-ordering Method}
\label{sec:TOM}
In this section we propose a systematic way of controlling the number
of species of lattice fermions within the requirement of
Nielsen-Ninomiya's no-go theorem \cite{CM}.  
We will first discuss the
$2$-dimensional case to give an intuitive understanding of this
mechanism.  Then we will go on to the $4$-dimensional case and show we
can construct a new minimally doubled action by this method.

Let us begin with the following simple $2$-dimensional Dirac operator
in momentum space with $O(a)$ Wilson-like terms.
\begin{align}
D(p)\,\,=\,\,  &( \sin p_{1} \,+\, \cos p_{1}\,-\,1)\, i \gamma_{1}
\nonumber\\
+ &( \sin p_{2} \,+\, \cos p_{2} \,-\, 1 )\, i \gamma_{2},
\label{2dUT}
\end{align}
where $p_{1}$, $p_{2}$ and $\gamma_{1}$, $\gamma_{2}$ stand for
$2$-dimensional momenta and Gamma matrices respectively.  Here the
deviation from the usual Wilson action is that the $O(a)$ terms are
accompanied by Gamma matrices.  There are four zeros of this Dirac
operator at 
$(\tilde{p}_{1},\tilde{p}_{2})=(0,0),(0,\pi/2),(\pi/2,0),(\pi/2,\pi/2)$.
This means that the number of the species is the same as for naive
fermions.

\begin{figure}[htbp]
\begin{minipage}{0.5\hsize}
\begin{center}
\includegraphics[height=5cm]{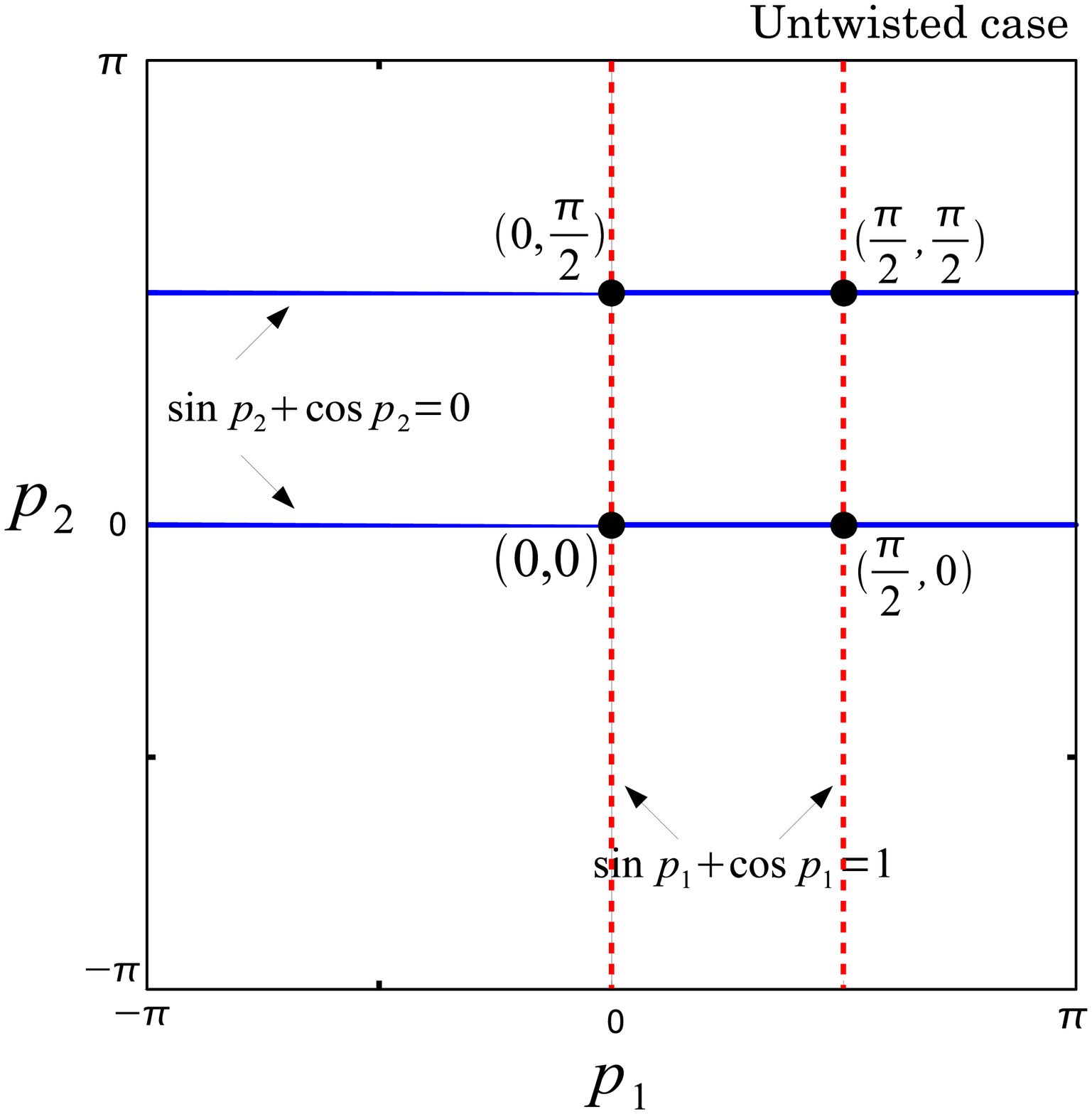} 
\end{center}
\caption{Red-dotted and blue-solid lines 
stand for zeros of
the coefficients of $\gamma_{1}$ and $\gamma_{2}$ respectively. 
Black points stand for the four zeros of the Dirac
operator.}
\label{UT}
\end{minipage}\,\,\,
\begin{minipage}{0,5\hsize}
\begin{center}
\includegraphics[height=5cm]{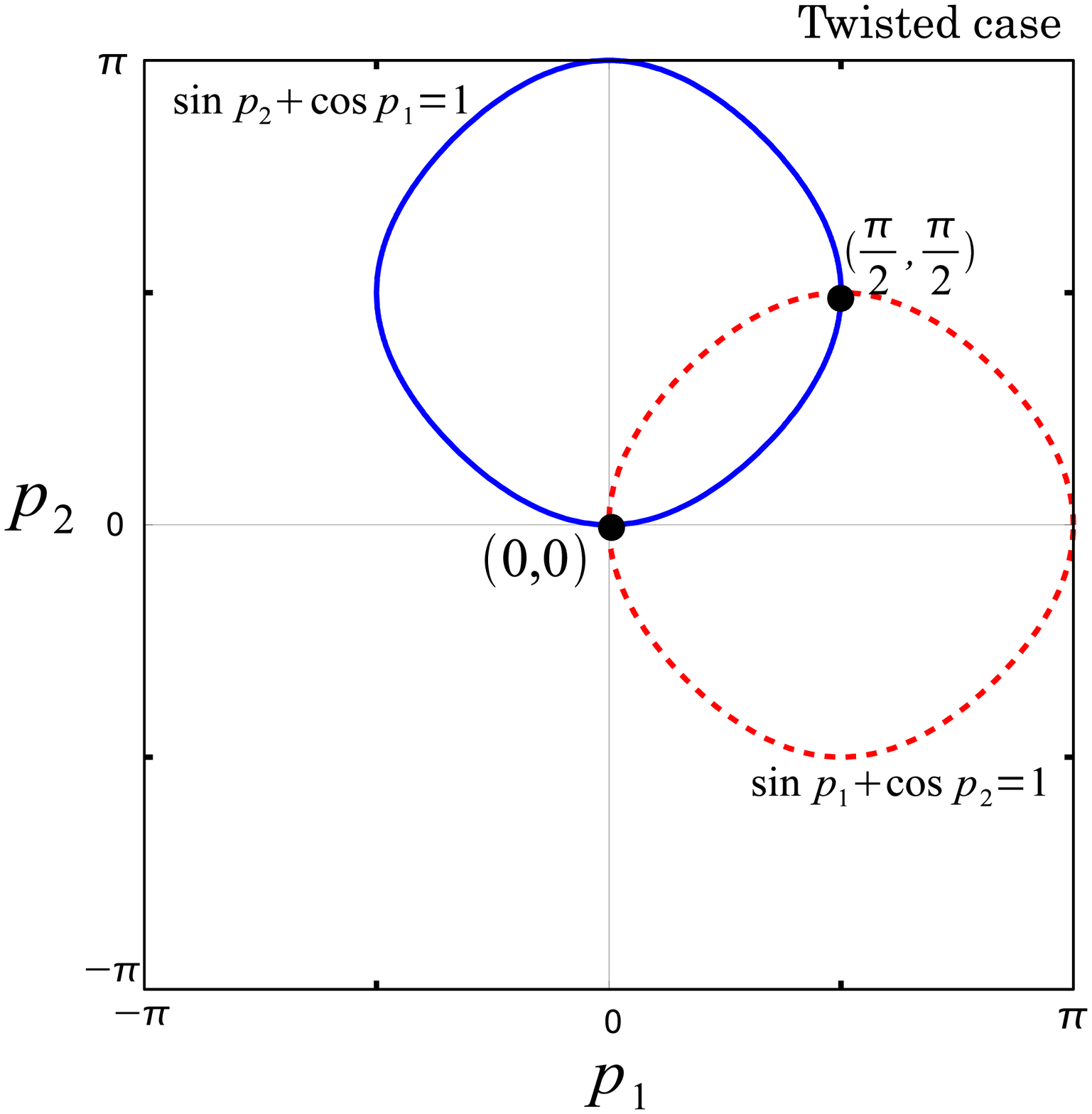} 
\end{center}
\caption{Red-dotted and blue-solid curves 
stand for zeros of the
coefficients of $\gamma_{1}$ and $\gamma_{2}$. 
Black points stand for two zeros ($0,0$)
($\pi/2,\pi/2$) of the Dirac operator.}
\label{1T}
\end{minipage}
\end{figure}

Next we ``twist'' the order of the $O(a)$ terms, or equivalently, permute
$\cos p_{1}$ and $\cos p_{2}$ to give
\begin{align}
D(p)\,\,=\,\,  &( \sin p_{1} \,+\, \cos p_{2}\,-\,1)\, i \gamma_{1}
\nonumber\\
+ &( \sin p_{2} \,+\, \cos p_{1} \,-\, 1 )\, i \gamma_{2}.
\label{2d1T}
\end{align}
Here only two of the zeros ($(0,0)$ and $(\pi/2,\pi/2)$) remain and
the other two are eliminated.  Thus the number of species becomes two,
the minimal number required by the no-go theorem.  As seen from this
example, twisting of the order of $O(a)$ terms reduces the number of
species.  We call this method ``twisted-ordering" in the rest of this
paper.  We depict the appearance of zeros in the Brillouin zone for
the above two cases in Fig.~\ref{UT} and \ref{1T}.

Now we consider excitations at the zeros.  
The Dirac operator is expanded about the zeros as
\begin{align}
D^{(1)}(q)\,\,&=\,\, i \gamma_{1}q_{1}  + \, i \gamma_{2}q_{2}\,+\, O(q^{2}),
\label{2d1T-Exp1}
\\
D^{(2)}(q)\,\,&=\,\, -i \gamma_{1}q_{2}  - \, i \gamma_{2}q_{1}\,+\, O(q^{2}).
\label{2d1T-Exp2}
\end{align}
Here we expand with respect to $q_{\mu}$ defined as
$p_{\mu}=\tilde{p}_{\mu}+q_{\mu}$ and we denote the two expansions as
$D^{(1)}$ for the zero ($0,0$) and $D^{(2)}$ for the other at
$(\pi/2,\pi/2)$.  Momentum bases at the zero $(0,0)$ are given by
${\bf b}^{(1)}_{1}=(1,0)$ and ${\bf b}^{(1)}_{2}=(0,1)$ while those at
the other $(\pi/2,\pi/2)$ are given by ${\bf b}^{(2)}_{1}=(0,-1) $ and
${\bf b}^{(2)}_{2}=(-1,0)$.  This means excitations from the two zeros
describe physical fermions on the orthogonal lattice.  
The gamma-5 at $\tilde{p}=(0,0)$ is given by
$\gamma_{5}^{(1)}\,=\,\gamma_{1}\gamma_{2}$
while the gamma-5 at $\tilde{p}=(\pi/2,\pi/2)$ is given by
$\gamma_{5}^{(2)}\,=\,\gamma_{2}\gamma_{1}\,=\,-\gamma_{5}^{(1)}$
This sign change between the species is a typical relation between two
species with minimal doubling since Nielsen-Ninomiya's no-go theorem 
requires fermion pairs to possess chiral charges with opposite signs.

We also obtain $4$d minimally doubled actions with this method.
The simplest form of 4d twisted-ordering actions is given by
\begin{align}
D(p)\,\,=\,\,  &( \sin p_{1} \,+\, \cos p_{2}\,-\,1)\, i \gamma_{1}
\nonumber\\
+ &( \sin p_{2} \,+\, \cos p_{3} \,-\, 1 )\, i \gamma_{2}
\nonumber\\
+ &( \sin p_{3} \,+\, \cos p_{4} \,-\, 1 )\, i \gamma_{3}
\nonumber\\
+ &( \sin p_{4} \,+\, \cos p_{1} \,-\, 1 )\, i \gamma_{4},
\label{4dMT}
\end{align}
with $O(a)$ Wilson-like terms whose order is twisted.
Here exist only $2$ zeros
\begin{equation}
(\tilde{p}_{1},\tilde{p}_{2},\tilde{p}_{3},\tilde{p}_{4})
=(0,0,0,0),\,\,\,(\pi/2,\pi/2,\pi/2,\pi/2).
\label{Z-4dMT}
\end{equation}
We have confirmed there are no other real zeros of this operator
numerically. This is a new type of minimally doubled fermion 
on the orthogonal lattice.
In this case excitations from the two zeros describe physical fermions 
on the orthogonal lattice again.  And the two associated $\gamma_{5}$'s
have opposite signs as
$\gamma_{5}^{(2)}\,=\,\gamma_{4}\gamma_{1}\gamma_{2}\gamma_{3}\,
=\,-\gamma_{5}^{(1)}$.

Gauging these theories is straightforward.  One merely inserts the gauge
fields as link operators in the hopping terms for the action in
position space.  Specifically, for the
twisted-ordering minimally doubled
fermion in position space we have
\begin{equation}
S\,\,=\,\, {1\over{2}}\sum_{n,\mu}\Big[ \bar{\psi}_{n}\,
  \gamma_{\mu}\, (U_{n,\mu}\psi_{n+\mu} -
  U^{\dag}_{n-\mu,\mu}\psi_{n-\mu}) \,+\, i \bar{\psi}_{n}\,
  \gamma_{\mu-1}\, (U_{n,\mu}\psi_{n+\mu}\, +\,
  U^{\dag}_{n-\mu,\mu}\psi_{n-\mu} \,-\, 2\psi_{n})\Big],
\label{ACTION}
\end{equation} 
where we define 
\begin{equation}
\mu-1 \equiv 
\left\{\begin{matrix}
		 1,2,3 & (\mu=2,3,4) \\
		4 & (\mu=1)				   
\end{matrix}\right.
\label{mu+1}
\end{equation}


\section{Minimally-Doubled Fermions}
\label{sec:MDF}

In this section we discuss two classes of minimally doubled actions
obtained from the original twisted-ordering action (\ref{4dMT}) \cite{CM}.  
The first one, which we call ``dropped twisted-ordering action'', is
constructed as following: We drop one of $\cos p_{\mu}-1$ terms in the
Dirac operator (\ref{4dMT}), for example, drop $\cos p_{1}-1$ as
\begin{align}
D(p)\,\,=\,\,  &( \sin p_{1} \,+\, \cos p_{2}\,-\,1)\, i \gamma_{1}
\nonumber\\
+ &( \sin p_{2} \,+\, \cos p_{3} \,-\, 1 )\, i \gamma_{2}
\nonumber\\
+ &( \sin p_{3} \,+\, \cos p_{4} \,-\, 1 )\, i \gamma_{3}
\nonumber\\
+ &  \sin p_{4}\, i \gamma_{4}.
\label{4dMT-D}
\end{align}
This has two zeros given by
\begin{equation}
(\tilde{p}_{1},\tilde{p}_{2},\tilde{p}_{3},\tilde{p}_{4})\,
=\,(0,0,0,0),\,\,\,\,\, (\pi,0,0,0). 
\label{Z-4dMT-D}
\end{equation}
It is obviously a minimally doubled action.

Next we consider a second variety
where we turn on a parameter in the action (\ref{4dMT}) as following,
\begin{align}
D(p)\,\,=\,\,  
  &( \sin p_{1} \,+\, \cos p_{2} \,-\, \alpha )\, i \gamma_{1}
\nonumber\\
+ &( \sin p_{2} \,+\, \cos p_{3} \,-\, \alpha )\, i \gamma_{2}
\nonumber\\
+ &( \sin p_{3} \,+\, \cos p_{4} \,-\, \alpha )\, i \gamma_{3}
\nonumber\\
+ &( \sin p_{4} \,+\, \cos p_{1} \,-\, \alpha )\, i \gamma_{4}.
\label{4dMT-a}
\end{align}
Here we replace unity with a positive parameter $\alpha$ in the
constant terms of the operator.  
The parameter range for minimal-doubling is given by
\begin{equation}  
0\,<\,\alpha \,<\, \sqrt{2}.
\label{MDR}
\end{equation}
In this range two zeros of the Dirac operator, 
which we denote as $\tilde{p}_{\mu}^{(1)}$
and $\tilde{p}_{\mu}^{(2)}$, are given by
\begin{align}
\tilde{p}_{\mu}^{(1)}\,&=\,
\arcsin \left({\alpha-\sqrt{2-\alpha^2}\over{2}} \right) 
\,=\, \arcsin A,  
\label{Z-4dMT-a1}
\\
\tilde{p}_{\mu}^{(2)}\,&=\,
\arcsin \left({\alpha+\sqrt{2-\alpha^2}\over{2}} \right) 
\,=\, \arcsin \sqrt{1-A^{2}}. 
\label{Z-4dMT-a2}
\end{align}
where for convenience we define a parameter $A$
as $A\equiv (\alpha-\sqrt{2-\alpha^2})/2$.
These two zeros reduce to Eq.~(\ref{Z-4dMT}) with $\alpha=1$.
By expanding the Dirac operator we see momentum bases are non-orthogonal 
for cases of $\alpha\not=1$;
thus, the associated lattices are non-orthogonal as with the
Creutz fermion\cite{BC}.


\section{Symmetries}
\label{sec:Sym}

In the previous section we have studied two classes of
minimally doubled actions in (\ref{4dMT-D}) and (\ref{4dMT-a}), both
of which are obtained from the original twisted-ordering action in (\ref{4dMT}).
In this section we will discuss discrete symmetries of these actions
and redundant operators generated by loop corrections.

Firstly we note both of the actions possess some common properties
with other minimally doubled fermions: ``Gamma-5 hermiticity", ``Discrete translation
invariance", ``flavor-singlet $U(1)_{V}$" and ``flavor-nonsinglet
$U(1)_{A}$." The last one is the exact chiral symmetry preventing
additive mass renormalization for the neutral pion.  
Besides them, they have exact
locality and gauge invariance when gauged by link variables.  We
expect every sensible minimally doubled fermion to possess these basic
properties.

On the other hand, Discrete symmetries associated with permutation of
the axes and $C$, $P$ or $T$ invariance depends on a class of the
actions.  In other words, we can identify minimally doubled actions by
these discrete symmetries.  Here we show discrete symmetries which the
dropped twisted-ordering action possesses: 1.$CP$, 2.$T$, 3.$Z_{2}$ 
associated with two zeros.
Here we take the $p_{4}$ direction as time.  
We also write the symmetries 
which the full twisted-ordering action possesses: 1.$CPT$, 2.$Z_{4}$ 
associated with axes permutation, 3.$Z_{2}$ associated with two zeros.
Because of lack of sufficient discrete symmetries,
redundant operators can be generated radiatively in
these action as in the other classes \cite{Bed}.  
For example, the case of the twisted-ordering action is similar to
that of Borici-Creutz action which has $S_{4}$ symmetry \cite{Bed}.  
We speculate the number of redundant 
operators generated in the twisted-ordering action
is the same as those of the Borici-Creutz one shown in \cite{Bed, Cap}.


\section{Classification}
\label{sec:CL}
In this section we discuss a classification of the known minimally
doubled fermions.  So far we have seen four variations on minimally
doubled fermions: Karsten-Wilczek \cite{KW}, Borici-Creutz \cite{BC}, twisted-ordering and
dropped twisted-ordering fermions \cite{CM}.  Here we classify these actions
into two types: One of them, which includes Karsten-Wilczek and
dropped twisted-ordering actions, is given by
\begin{equation} 
D(p)\,=\, \sum_{\mu}i\gamma_{\mu}\sin p_{\mu}\,
+\, \sum_{i,j} i\gamma_{i}R_{ij}(\cos p_{j}-1),
\label{GENE1}
\end{equation}
where we take $i=1,2,3$ as $i$ while $j=2,3,4$ as $j$. 
The point is that indices $i$ and $j$
are staggered.  
The different
actions in this class depend on the choice for the matrix $R$.
For example, consider the following $R$'s
\begin{equation}
R=\left( 
\begin{matrix}
1\,\, & 1\,\, & 1 \\
0\,\, & 0\,\, & 0 \\
0\,\, & 0\,\, & 0 \\
\end{matrix}
\right),\,\,\,\,\,\,\,
\left( 
\begin{matrix}
1\,\, & 0\,\, & 0 \\
0\,\, & 1\,\, & 0 \\
0\,\, & 0\,\, & 1 \\
\end{matrix}
\right),\,\,\,\,\,\,\,
\left( 
\begin{matrix}
1\,\, & 1\,\, & 0 \\
0\,\, & 0\,\, & 1 \\
0\,\, & 0\,\, & 0 \\
\end{matrix}
\right).
\label{R}
\end{equation}
For the first $R$ the general form (\ref{GENE1}) reduces to the
Karsten-Wilczek action while it reduces to the dropped
twisted-ordering action for the second one.
The action for the third case is a new possibility.

We see that there are many options associated with possible $R$'s.
One common property with fermions obtained in this way is that two
zeros are separated along a single lattice axis and given by
$(0,0,0,0)$ and $(\pi,0,0,0)$.  One significance about the general
form (\ref{GENE1}) is that a coefficient of at least one Gamma matrix
has no $(\cos p_{\mu}-1)$ term.  It is also notable that, in a
coefficient of each gamma matrix, a momentum component $p_{\mu}$
associated with a $\sin$ term differs from the component in any $\cos$
term.  These two points seem to be essential to minimal doubling.

The second class of actions includes the twisted-ordering fermion with
the $\alpha$ parameter and the Borici-Creutz actions.  A generalized
Dirac operator for this type is given by
\begin{equation}
D(p)\,\,=\,\, i\sum_{\mu}[ \gamma_{\mu}\sin (p_{\mu}+\beta_{\mu}) \,
-\, \gamma_{\mu}'\sin (p_{\mu}-\beta_{\mu}) ]\,-\, i\Gamma
\label{GENE2}
\end{equation} 
where $\gamma_{\mu}'=A_{\mu\nu}\gamma_{\nu}$ is another set of gamma
matrices where we define $A$ as an orthogonal matrix with some
conditions: At least one eigenvalue of $A$ should be $1$ and all four
components of the associated eigenvector should have non-zero values.
Here $\beta_{\mu}$ and $\Gamma$ has a relation with this $A$ as
$\Gamma=\sum_{\mu}\gamma_{\mu}\sin
2\beta_{\mu}=\sum_{\mu}\gamma_{\mu}'\sin 2\beta_{\mu}$, which means
$\sin 2\beta_{\mu}$ is an eigenvector of $A$ as $A_{\mu\nu}\sin
2\beta_{\nu}=\sin 2\beta_{\mu}$.  Thus once $A$ is fixed,
$\beta_{\mu}$ and $\Gamma$ are determined up to a overall factor of
$\sin 2\beta_{\mu}$.  By imposing these conditions on $A$,
$\beta_{\mu}$ and $\Gamma$, the action (\ref{GENE2}) can be a
minimally doubled action.  In such a case $\beta_{\mu}$ indicates
locations of two zeros as $\tilde{p}_{\mu}=\pm\beta_{\mu}$.  Adjusting
$\beta$, we can control the locations of two zeros.
Note that the first and second terms in (\ref{GENE2}) are nothing but
naive fermion actions.  We can eliminate some species by combining two
naive actions with different zeros in one action.

Now let us show this action includes minimally doubled actions.
(\ref{GENE2}) reduces to the Borici action and the twisted-ordering action by
choosing $A$'s as
\begin{align} 
\gamma'_{\mu} \,&=\, A_{\mu\nu} \gamma_{\nu},
\nonumber\\
A \,&=\,  {1\over{2}}\left( 
\begin{matrix}
-1 & 1 & 1 & 1 \\
1 & -1 & 1 & 1 \\
1 & 1 & -1 & 1 \\
1 & 1 & 1 & -1
\end{matrix}
\right) 
\,\,\,\,\,\,\,{\rm and}\,\,\,\,\,\,\,
\left( 
\begin{matrix}
0 & 0 & 0 & 1 \\
1 & 0 & 0 & 0 \\
0 & 1 & 0 & 0 \\
0 & 0 & 1 & 0
\end{matrix}
\right), 
\label{GENE2-T}
\end{align}
where we fix $\beta_{\mu}=\pi/4$ and $\Gamma$ is given by
$\Gamma=\sum_{\mu}\gamma_{\mu}$.
Finally we have classified all the known minimally doubled fermions
into the two types.  Now we can derive a lot of varieties from these
general forms deductively as we have shown in (\ref{R}).

\section{Higher-dimensional Extensions}

In this section we discuss higher-dimensional extensions of minimally doubled actions.
As two of the present authors showed in \cite{KM}, we can extend minimally doubled fermions
to general dimensions.
The main observation in this generalization is that the parameter range for minimal-doubling 
tends to get narrower in higher dimensions. 
For example, one of $d$-dimensional generalizations of Karsten-Wilczek fermion is given by
\begin{equation}
D_{KW}=\sum_{\mu}^{d}[i\gamma_{\mu}\sin p_{\mu}\,+\,i\gamma_{d}(r-\cos p_{\mu})],
\end{equation}
where $r$ is a parameter which will be fixed to realize minimal-doubling.
Here the parameter range for minimal-doubling is given by
\begin{equation}
{d+\sqrt{2}-3\over{d}}\,<\,r\,<\,{d+\sqrt{2}-1\over{d}}.
\end{equation}
It is obvious that this range gets narrower with the dimension ($d$) getting larger.
This is also the case with even-dimensional generalization of Borici-Creutz action, 
which has two parameters $B$ and $C$.
The minimal-doubling range for $C$ is given by
\begin{equation}
{d-2\over{d}}\,<\,C\,<\,1
\end{equation}
where we concentrate only on the case $d=2m$ ($m=1,2,3...$).
Here this range gets narrower with the dimension.
As seen from these two examples, it is more difficult to realize minimal-doubling in higher dimensions.
It is natural since the number of species increases with the dimension and 
we need to impose a stronger condition to obtain only two species in higher dimensions.
On the other hand things are a little different in the twisted-ordering fermion.
In this case the minimal-doubling parameter range for $\alpha$ in Eq.~(\ref{MDR}) 
holds even in higher dimensions \cite{CM}.
This is because the mechanism ``Twisted-ordering" itself reduces the number of species 
in this case while adjusting the parameter controls them in the other cases.
This fact may imply the twisted-ordering fermion is the most fundamental type
among all the known minimally-doubled fermions.

\section{Conclusion}
In this paper we propose a method to control 
the number of species of lattice fermions and 
obtain new classes of minimally-doubled fermions. 
We also classify all the known minimally doubled actions into two types.  
By this classification we can derive several further minimally
doubled actions deductively.
We also study a higher-dimensional extension of them and find 
a parameter range for minimal-doubling gets narrower with the dimension 
in most cases except for the twisted-ordering fermion.     
With several varieties of minimally doubled fermion actions available,
a goal is to apply these actions to numerical simulations and study
their relative advantages.
Such studies \cite{Cap} so far focused only on Karsten-Wilczek and Borici-Creutz fermions.  
One can now explore other varieties such as ones in this paper.


\begin{thebibliography}{9}

\bibitem{KW}
L.~H.~Karsten, Phys. Lett. B {\bf 104}, 315 (1981);
F.~Wilczek, Phys. Rev. Lett. {\bf 59}, 2397 (1987). 

\bibitem{BC}
M.~Creutz, JHEP 0804, 017 (2008);
A.~Borici, Phys. Rev. D {\bf 78}, 074504 (2008).

\bibitem{CM}
M.~Creutz and T.~Misumi, Phys. Rev. D {\bf 82}, 074502 (2010) [arXiv:1007.3328 [hep-lat]].

\bibitem{KM}
T.~Kimura and T.~Misumi, Prog. Theor. Phys. {\bf 124}, 415 (2010) [arXiv:0907.1371 [hep-lat]];
Prog. Theor. Phys. {\bf 123}, 63 (2010) [arXiv:0907.3774 [hep-lat]].

\bibitem{Bed}
P.~F.~Bedaque, M.~I.~Buchoff, B.~C.~Tiburzi and A.~Walker-Loud, Phys. Lett. B {\bf 662}, 449 (2008) [arXiv:0801.3361 [hep-lat]];
Phys. Rev. D {\bf 78}, 017502 (2008) [arXiv:0804.1145 [hep-lat]]. 

\bibitem{Cap}
S.~Capitani, J.~Weber, H.~Wittig, Phys. Lett. B {\bf 681}, 105 (2009) [arXiv:0907.2825 [hep-lat]];
S.~Capitani, M.~Creutz, J.~Weber, H.~Wittig, JHEP 1009, 027 (2010) [arXiv:1006.2009 [hep-lat]].



\end{thebibliography}
\end{document}